\documentclass[A4,aps,onecolumn,superscriptaddress]{revtex4}
\usepackage[dvips]{graphics}
\usepackage{enumerate}
\usepackage{amsmath}
\usepackage{amssymb}
\usepackage{amsfonts}
\usepackage{latexsym}
\usepackage{revsymb}
\usepackage{bm}

\begin{document}
\begin{flushright}
VPI-IPPAP-06-06
\end{flushright}

\title{Constraints on New Physics from Matter Effects on Neutrino Oscillation}
\author{Minako~Honda}
\affiliation{Physics Department, Ochanomizu Women's University, Tokyo 112-8610, Japan}
\author{Yee~Kao}
\affiliation{Physics Department, Virginia Tech, Blacksburg VA 24061, USA}
\author{Naotoshi~Okamura}
\affiliation{Yukawa Institute for Theoretical Physics, Kyoto University, Kyoto 606-8502, Japan}
\author{Alexey~Pronin}
\affiliation{Physics Department, Virginia Tech, Blacksburg VA 24061, USA}
\author{Tatsu~Takeuchi}
\affiliation{Physics Department, Virginia Tech, Blacksburg VA 24061, USA}

\maketitle

\noindent
When considering matter effects on neutrino oscillation, it is customary to consider only the 
$W$-exchange interaction of the $\nu_e$ with the electrons in matter. However, if new physics that distinguishes among the three generations of neutrinos exist, it can lead to extra matter effects via 
radiative corrections to the $Z\nu\nu$ vertex which violates neutral current universality, or via the direct exchange of new particles between the neutrinos and the matter particles.
To consider this possibility, we analyze neutrino oscillation in matter 
governed by the effective Hamiltonian \cite{HKOT}
\begin{equation}
H = 
\tilde{U}
\left[ \begin{array}{ccc} \lambda_1 & 0 & 0 \\
                          0 & \lambda_2 & 0 \\
                          0 & 0 & \lambda_3
       \end{array}
\right]
\tilde{U}^\dagger
= U
\left[ \begin{array}{ccc} 0 & 0 & 0 \\
                          0 & \delta m^2_{21} & 0 \\
                          0 & 0 & \delta m^2_{31}
       \end{array}
\right]
U^\dagger + a
\left[ \begin{array}{ccc} 1 & 0 & 0 \\
                          0 & -\xi/2 & 0 \\
                          0 & 0 & \xi/2 
       \end{array}
\right] \;,
\end{equation}
where $U$ is the MNS matrix, and $a=2EV_{CC}=2\sqrt{2}E G_F N_e$.  
$\xi$ parametrizes the contribution of new physics.

The extra contribution can manifest itself when $a>|\delta m^2_{31}|$ (\textit{i.e.} $E\agt 10$~GeV for typical matter densities in the Earth) in the $\nu_\mu$ survival probability as \cite{HKOT}
\begin{equation}
P(\nu_\mu\rightarrow\nu_\mu) 
\approx 1-\sin^2\left(2\theta_{23} - \frac{a\xi}{\delta m^2_{31}}\right)\sin^2\dfrac{\tilde{\Delta}}{2}\;,\qquad
\tilde{\Delta} \approx \Delta_{31} c_{13}^2 - \Delta_{21} c_{12}^2\;.
\end{equation}
(This expression assumes that the CP violating phase $\delta$ is zero.)
If $\sin^2(2\theta_{23})=1$, then the small shift due to $\xi$ will be invisible.  
However, if $\sin^2(2\theta_{23})=0.92$ (the current 90\% lower bound), 
and if $\xi$ is as large as $\xi=0.025$ (the 
central value from CHARM \cite{CHARM}), then the shift in the survival probability at the
first oscillation dip can be as large as $\sim40\%$.
If the NUMI beam at Fermilab in its high-energy mode \cite{NUMI} were aimed at a declination angle of
$46^\circ$ toward a Mega-ton class detector at Kamioka \cite{HyperK} (baseline 9120~km), 
such a shift would be visible after just one year of data taking (assuming a Mega-ton fiducial volume and
100\% efficiency).  
The absence of any shift after 5 years of data taking would constrain $\xi$ to $0 \pm 0.005$ \cite{HKOT}.

A constraint on $\xi$ at that level can potentially place strong constraints on possible
new physics. For instance, the lower bound on the mass of a $Z'$ which couples to
$B-3 L_\tau$ \cite{MA} will be $\sim 5$~TeV if we assume its coupling to be comparable to the
$SU(2)_L$ coupling $g\approx 0.65$. In contrast, the lower bound from currently available
LEP and SLD data is only $\sim 600$~GeV \cite{CLLT}.
Details of this analysis will be presented in Ref.~\cite{HKOPT}. There, constraints on 
generation-non-diagonal leptoquarks, 
R-parity violating SUSY models, extended Higgs models, etc. will be discussed.

\section*{Acknowledgments}

This paper was presented as a poster by Honda, Okamura, and Takeuchi at the 
YITP workshop `Progress in Particle Physics',  July 31, 2006.
This research was supported in part (A.P. \& T.T.) by the U.S. Department of Energy,
grant DE-FG05-92ER40709, Task A.



\begin{thebibliography}{99}

\bibitem{HKOT}
M.~Honda, Y.~Kao, N.~Okamura, and T.~Takeuchi, hep-ph/0602115;\\
M.~Honda, N.~Okamura, and T.~Takeuchi, hep-ph/0603268.

\bibitem{CHARM}
  J.~Dorenbosch {\it et al.}  [CHARM Collaboration],
  Phys.\ Lett.\ B {\bf 180}, 303 (1986);\\
  P.~Vilain {\it et al.}  [CHARM-II Collaboration],
  Phys.\ Lett.\ B {\bf 320}, 203 (1994).

\bibitem{NUMI}
NUMI Technical Design Handbook, 
{\sf http://www-numi.fnal.gov/numwork/tdh/tdh\_index.html}


\bibitem{HyperK}
  Y.~Itow {\it et al.},
  arXiv:hep-ex/0106019;
updated version available at {\sf http://neutrino.kek.jp/jhfnu/}.


\bibitem{MA}
  E.~Ma,
  Phys.\ Lett.\ B {\bf 433}, 74 (1998)
  [arXiv:hep-ph/9709474];
  E.~Ma and D.~P.~Roy,
  Phys.\ Rev.\ D {\bf 58}, 095005 (1998)
  [arXiv:hep-ph/9806210];
  E.~Ma and U.~Sarkar,
  Phys.\ Lett.\ B {\bf 439}, 95 (1998)
  [arXiv:hep-ph/9807307].

\bibitem{CLLT}
  L.~N.~Chang, O.~Lebedev, W.~Loinaz and T.~Takeuchi,
  Phys.\ Rev.\ D {\bf 63}, 074013 (2001)
  [arXiv:hep-ph/0010118].

\bibitem{HKOPT}
M.~Honda, Y.~Kao, N.~Okamura, A.~Pronin, and T.~Takeuchi, in preparation.


\end{thebibliography}
\end{document}